\documentclass[aps,prl,groupedaddress,showpacs,twocolumn,floatfix]{revtex4}

\usepackage{graphicx}
\usepackage{longtable}

\def\gtorder{\mathrel{\raise.3ex\hbox{$>$}\mkern-14mu
 \lower0.6ex\hbox{$\sim$}}}
\def\ltorder{\mathrel{\raise.3ex\hbox{$<$}\mkern-14mu
 \lower0.6ex\hbox{$\sim$}}}
\def\mugegm{\mu_p G_E / G_M}
\def\gegm{G_E / G_M}
\def\ge{G_E}
\def\gm{G_M}
\def\gd{G_D}
\def\gegmtilde{\widetilde{G}_E / \widetilde{G}_M}
\def\getilde{\widetilde{G}_E}
\def\gmtilde{\widetilde{G}_M}
\def\f3{\widetilde{F}_3}
\def\y2g{Y_{2\gamma}}
\def\etal{\textit{et al.}}

\begin{document}

\title{Extraction of two-photon contributions to the proton form factors}

\author{J. Arrington}

\affiliation{Physics Division, Argonne National Laboratory, Argonne, Illinois 60439, USA}

\date{\today}

\begin{abstract}

Significant discrepancies have been observed between proton form factors as
measured by Rosenbluth separation and polarization transfer techniques.  There
are indications that this difference may be caused by corrections to the one
photon exchange approximation that are not taken into account in standard
radiative correction procedures.  In this paper, we constrain the two-photon
amplitudes by combining data from Rosenbluth, polarization transfer, and
positron-proton scattering measurements.  This allows a rough extraction of
these two-photon effects in elastic electron-proton scattering, and provides
an improved extraction of the proton electromagnetic form factors.

\end{abstract}

\maketitle


\section{Introduction}

Extractions of the proton electromagnetic form factors utilizing the
polarization transfer technique show a significant decrease in the ratio of
electric to magnetic form factor at large momentum transfers~\cite{jones00,
gayou02}.  These results contradict a large body of Rosenbluth separation
measurements~\cite{arrington03a, christy04, arrington03e} which indicate
approximate scaling of the form factors, $\mugegm \approx 1$.  This
inconsistency leads to a large uncertainty in our knowledge of the proton
electromagnetic form factors and could have significant implications for
other experiments which rely on similar techniques or which assume knowledge
of the proton form factors to interpret their data~\cite{budd03, dutta03,
arrington04a}.

It has been suggested that two-photon exchange contributions could
be responsible for the discrepancy between the Rosenbluth, or
Longitudinal-Transverse (L-T), separation and polarization transfer form
factors~\cite{guichon03}.  Calculations of the two-photon
exchange diagram suggest that this may indeed be the case~\cite{blunden03,
chen04}, and there is some evidence for two-photon exchange in comparisons of
electron-proton and positron-proton scattering~\cite{arrington04b}. However, a
complete calculation of two-photon exchange must include contributions where
the intermediate proton is in an excited state, which are not included in
Ref.~\cite{blunden03}.  In Ref.~\cite{chen04}, the contribution from
intermediate states is included through two-photon scattering off of partons
in the proton, with emission and re-absorption of the partons by the nucleon
described in terms of generalized parton distributions. However, this approach
is not expected to be valid at low four-momentum transfers, $Q^2$, or for
small values of the virtual photon polarization, $\varepsilon$, and
yields approximately half of the effect needed to bring the two techniques
into agreement at larger $Q^2$.

A general formalism does exist for parameterizing contributions
beyond the one-photon (Born) approximations~\cite{guichon03}. While discussed
in terms of two-photon contributions, this formalism includes all terms in the
elastic scattering amplitude: vertex corrections, loop corrections (vacuum
polarization), soft and hard two-photon contributions, and multi-photon
exchange; all terms with just the electron and proton in the final state.  In
the Born approximation, one obtains two real amplitudes which depend only on
the momentum transfer: $\ge(Q^2)$ and $\gm(Q^2)$.  In the generalized case,
there are three complex amplitudes which depend on both $Q^2$ and
$\varepsilon$: $\getilde(\varepsilon,Q^2)$, $\gmtilde(\varepsilon,Q^2)$, and
$\f3(\varepsilon,Q^2)$. For convenience, we break up the generalized form
factors into the Born values and the ``two-photon'' contributions, e.g.
$\getilde(\varepsilon,Q^2) = \ge(Q^2) + \Delta\ge(\varepsilon,Q^2)$, and
define $\y2g$,
\begin{equation}
\y2g = {\cal R}e \biggl( \frac{\nu\f3}{M_p^2 \mid \gm \mid} \biggr),
\end{equation}
where $\nu = M_p^2\sqrt{(1+\varepsilon)/(1-\varepsilon)}\sqrt{\tau(1+\tau)}$
(equivalent to the definition given in Ref.~\cite{guichon03}).  We now have
the two usual Born-level form factors and three two-photon amplitudes:
$\Delta\ge$, $\Delta\gm$, and $\y2g$. The first two are complex, but as long
as they are not too large, only the real portion of these amplitudes has any
significant effect on the observables discussed below, so throughout this
paper we will refer only to the real part of $\getilde$ and $\gmtilde$.

The goal of this work is to use the existing data on elastic electron-proton
and positron-proton scattering data to estimate the small two-photon
amplitudes, and then use these amplitudes to correct the form factors
extracted from polarization transfer and Rosenbluth separation measurements.

\section{Extraction of the two-photon amplitudes}
The proton form factor ratio, $\gegm$, has been extracted from cross section
and polarization transfer measurements assuming one-photon exchange.  In the
generalized formalism, the extracted ratio does not yield the true form
factor ratio, but is a function of these generalized form factors:
\begin{eqnarray}
\label{eq:r_poltrans}
R_{Pol} = (\gegmtilde) +
\bigl ( 1 - \frac{2\varepsilon}{1+\varepsilon} \gegmtilde) \y2g, \\
\label{eq:r_rosenbluth}
R_{L-T}^2 = (\gegmtilde)^2 +
2 ( \tau + \gegmtilde ) \y2g,
\end{eqnarray}
where $\tau = Q^2/4M_p^2$. Keeping terms up to order $\alpha_{EM}$, the change to the
reduced cross section ($\sigma_r = \tau \gm^2 + \varepsilon \ge^2$ in the Born
approximation) is
\begin{equation}
\label{eq:delta_sigma}
\frac{\Delta\sigma_r}{\gm^2} \approx 2 \tau \frac{\Delta\gm}{\gm}
+ 2 \varepsilon \rho^2 \frac{\Delta\ge}{\ge}
+ 2 \varepsilon (\tau+\rho) \y2g
\end{equation}
where $\rho = \ge/\gm$.

The general procedure for extracting the two photon amplitudes is as follows:
From Eqs.~\ref{eq:r_poltrans} and \ref{eq:r_rosenbluth}, we can see that it is
only the $\y2g$ term that leads to a difference between the polarization
transfer and L-T form factor ratio, and so this difference will allow us to
determine $\y2g$. To obtain the true (Born) form factors we must still
determine $\Delta\gm$ and $\Delta\ge$.  Because the the dominant terms of the
two-photon correction changes sign for positron-proton scattering, we can use
the existing data for positron-proton scattering as an additional constraint
on $\Delta\ge$ and $\Delta\gm$, allowing an extraction of the true form
factors, $\ge$ and $\gm$, corrected for two-photon (and multi-photon) exchange
contributions.  These are the form factors that can be directly connected to
the structure of the proton, and which can be compared to lattice calculations
or models of the nucleon.

Given the limitations of the existing cross section, polarization transfer,
and positron-proton scattering data, we are forced to make some assumptions
in the extraction of the two-photon amplitudes.  First, we assume that
two-photon effects are responsible for all of the discrepancy. Second, we
assume that the two-photon amplitudes depend weakly on $\varepsilon$, although
we will examine the effect of $\varepsilon$-dependence in the error analysis.
Finally, we only consider processes that are of order $\alpha_{EM}$ with
respect to the born amplitudes, and neglect higher order corrections.
Specifically, we neglect terms other than the ``standard'' radiative
corrections, two-photon exchange, and soft multi-photon exchange (``Coulomb
distortion'') which is ${\cal O}(\alpha_{EM})$ after resummation.  With
these assumptions, it is possible to constrain the ``two-photon''
contributions to the form factors well enough to extract $\ge$ and $\gm$,
albeit with additional uncertainty due to these two-photon corrections. 
Further data would allow improved extractions of the two-photon amplitudes, as
well as provide better tests of the assumptions used in the analysis.

In the analysis of the existing cross section data, two-photon exchange
contributions were only treated approximately, while other radiative
corrections of ${\cal O}(\alpha_{EM})$ were applied. However, except for the
well understood Bremsstrahlung corrections, all of these terms are included
in the two-photon amplitudes $\Delta\ge$ and $\Delta\gm$.  In principle, the
corrections that were applied to the measured cross sections should be removed
before using the data to extract these amplitudes. However, it turns out that
this is not necessary. The extraction of $\y2g$ comes from the difference
between $R_{Pol}$ and $R_{L-T}$, but $R_{Pol}$ is insensitive to these
radiative corrections, while $R_{L-T}$ depends only on the
\textit{$\varepsilon$-dependent} corrections to the cross section. Because the
loop and vertex corrections are independent of $\varepsilon$, they do not
modify the value of $R_{L-T}$ extracted from the data, and so do not change
the extracted value of $\y2g$.

The other two-photon amplitudes will be constrained by the comparison of
electron-proton and positron-proton change sign with the charge of the lepton
are extracted: the two-photon exchange and the Coulomb corrections. Because
the loop and vertex corrections are identical for positron and electron
scattering, their inclusion does not modify the comparison of positron and
electron data.  Strictly speaking, the contributions due to loop and vertex
diagrams are not \textit{corrections} to the generalized form factors, they
are included in $\Delta\gm$ and $\Delta\ge$. However, the goal is to obtain
the true form factors and applying the loop and vertex to the measured cross
sections yields the same result as including them in the two-photon
corrections $\Delta\ge$ and $\Delta\gm$. Similarly, while soft multi-photon
exchange (Coulomb distortion) can play a non-negligible role at
low-to-moderate $Q^2$ values~\cite{arrington04c}, these corrections should not
be applied to the data for this analysis, as they are included as part of the
higher-order corrections ($\Delta\gm$, $\Delta\ge$, and $\y2g$).

\begin{figure}[htb]
\includegraphics[height=5.5cm,width=8.0cm,angle=0]{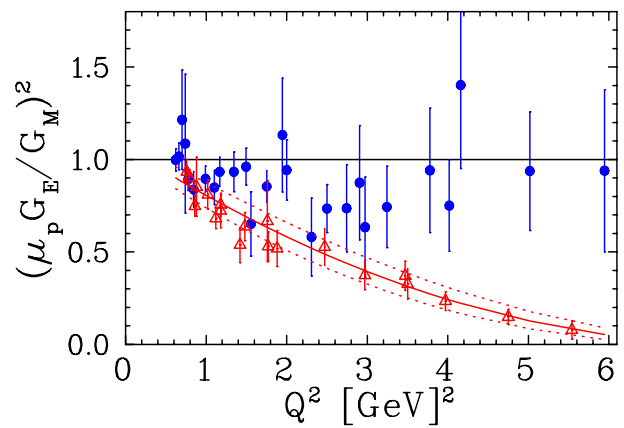}
\caption{(Color online) Rosenbluth form factor ratio squared, $R^2_{L-T}$ (blue
circles), polarization transfer ratio squared, $R^2_{Pol}$ (red triangles),
and the parameterization of the polarization transfer ratio and uncertainty
(solid and dotted red lines) used in the extraction of $\y2g$.  The 
error bars shown on the Rosenbluth extractions include an estimate of the
uncertainty in the determination of the normalization of the different data
sets in the global analysis.
\label{fig:ros_pol}}
\end{figure}

To extract $\y2g$, we compare the Rosenbluth extraction of $\gegm$ from a
global analysis of cross section data (Fig. 2 of Ref.~\cite{arrington04a}) to
a parameterization of the polarization transfer results and uncertainties, as
shown by solid and dotted lines in Fig.~\ref{fig:ros_pol}.  We limit ourselves
to $0.6 < Q^2 < 6.0$~GeV$^2$, to match the $Q^2$ range of precise polarization
transfer data. We then use the difference between $R_{L-T}$ and $R_{Pol}$
to determine $\y2g$, using $R_{Pol}$ as the approximate value for $\gegmtilde$
in Eqs.~\ref{eq:r_poltrans} and \ref{eq:r_rosenbluth}.  Note that if one
instead uses the final value of $\gegmtilde$ extracted from this analysis, the
change is negligible. There is no way to extract the $\varepsilon$-dependence
of $\y2g$ because we only have a single value of $R_{L-T}$ at each $Q^2$
value, taken from the full $\varepsilon$ range of the cross section data, and
because the polarization transfer ratios have not been measured at different
$\varepsilon$ values. However, if the $\varepsilon$-dependence in the
amplitudes is large enough to introduce a significant nonlinearity, then the
reduced cross section as a function of $\varepsilon$ would deviate from the
linearity predicted in the one-photon exchange approximation.  While current
data is not precise enough to set tight limits on deviations from linearity,
the existing data is all consistent with a linear dependence. Therefore, we
assume that $\getilde$, $\gmtilde$, and $\y2g$ are independent of
$\varepsilon$.  Of course, given a model of the $\varepsilon$-dependence, or
measurements of the nonlinearities in the two-photon exchange effects, we
could incorporate this information on the $\varepsilon$-dependence into the
fit to make an improved extraction.

\begin{figure}[htb]
\includegraphics[height=5.5cm,width=8.0cm,angle=0]{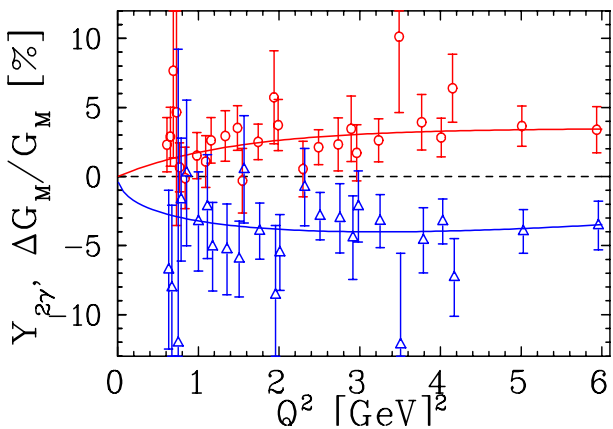}
\caption{(Color online) Extracted values of $\y2g$ (red circles) and
$\Delta\gm/\gm$ (blue triangles), along with fits to the extracted amplitudes.
The fits and parameterized uncertainties are given in the appendix.
\label{fig:amplitudes}}
\end{figure}

Figure~\ref{fig:amplitudes} shows the extracted values for $\y2g$ as a
function of $Q^2$, along with a fit to these extracted values, given in
Eq.~\ref{eq:fit_y2g}. Guichon and Vanderhaeghen~\cite{guichon03} performed a
similar extraction of $\y2g$, using fits to $R_{Pol}$ and $R_{L-T}$. They
extracted $\gegm$ from the polarization transfer data, correcting for the
contribution of $\y2g$ and assuming that $\Delta\ge$ and $\Delta\gm$ are
negligible. However, while $\y2g$ yields the \textit{difference} between the
two techniques, as well as the largest correction to the recoil polarization
ratio, comparisons of positron-proton and electron-proton scattering
demonstrate that $\Delta\gm$ and $\Delta\ge$ cannot be
neglected~\cite{arrington04b}. The values of $\y2g$ required to explain the
discrepancy yield a 5--8\% enhancement of the electron cross section at large
$\varepsilon$. Because the dominant term of the two-photon correction changes
sign for positron-proton scattering, this would imply a decrease in the
positron-proton cross section, and a ratio of positron to electron scattering
of $\ltorder$0.9. Data from Mar \etal~\cite{mar68} at large $\varepsilon$ and
$Q^2$ yields an average ratio of 1.017$\pm$0.024, well above this expectation.

One therefore needs additional input to constrain $\Delta\gm$ and $\Delta\ge$.
Precise comparisons of positron and electron scattering over a wide range in
$Q^2$ and $\varepsilon$ would allow the extraction of these amplitudes, but the
positron-proton scattering data above $Q^2=1.3$~GeV$^2$ is limited to small
scattering angles, corresponding to $\varepsilon>0.7$. Because of the very
limited $\varepsilon$ range, the positron data cannot be used to constrain the
\textit{$\varepsilon$-dependence} at large $Q^2$ values. However, they still
provide a useful constraint for the two-photon amplitudes. The positron data
at large $\varepsilon$ all indicate small two-photon contributions. To be
consistent with this data, the contribution of $\y2g$ to the cross section at
large $\varepsilon$ must be cancelled by the contributions of $\Delta\ge$ and
$\Delta\gm$. The change in the cross section due to $\Delta\ge/\ge$ is
suppressed with respect to the contribution from $\Delta\gm/\gm$ by a factor
of $\varepsilon\rho^2/\tau$ (Eq.~\ref{eq:delta_sigma}), which is below 0.15
for $Q^2$=2~GeV$^2$ and below 0.01 for $Q^2=5.6$~GeV$^2$. Therefore, unless
$\Delta\ge/\ge$ is much larger than $\Delta\gm/\gm$, the $\y2g$ contribution
to the cross section at large $\varepsilon$ must be cancelled almost entirely
by $\Delta\gm$.  Given the value of $\y2g$, we can determine $\Delta\gm$ by
requiring that the two-photon contribution to the cross section
(Eq.~\ref{eq:delta_sigma}) from $\y2g$ at $\varepsilon=1$ be cancelled by the
contribution from $\Delta\gm$: $\Delta\gm/\gm = - (1 + \rho/\tau) \y2g$.
Figure~\ref{fig:amplitudes} shows $\Delta\gm/\gm$ as determined from the
above procedure, as well as a fit to these extracted values
(Eq.~\ref{eq:fit_dgmgm}).  Note that these amplitudes are a few percent of the
Born amplitudes, larger than previously believed but still of order
$\alpha_{EM}$.

The remaining two-photon amplitude, $\Delta\ge$, is more difficult to
constrain, as it has a much smaller effect on the cross section.  However,
because both $\y2g$ and $\Delta\ge$ yield a correction to the cross section
that is proportional to $\varepsilon$, the $\varepsilon \rightarrow 0$ limit
can be used to extract $\gm$ with minimal uncertainty from $\y2g$ and
$\Delta\ge$.  So the lack of information on $\Delta\ge$ only affects extracted
values of $\gegm$. For this analysis, we take $\Delta\ge$ to be
zero, and use the larger of $\y2g$ and $\Delta\gm/\gm$ as an estimate of the
uncertainty of $\Delta\ge$.  This yields an additional uncertainty on the
extracted value of $\gegm$ of approximately 3--4\%, which is smaller than the
typical experimental uncertainties from the polarization transfer data.

\begin{figure}[htb]
\includegraphics[height=5.5cm,width=8.0cm,angle=0]{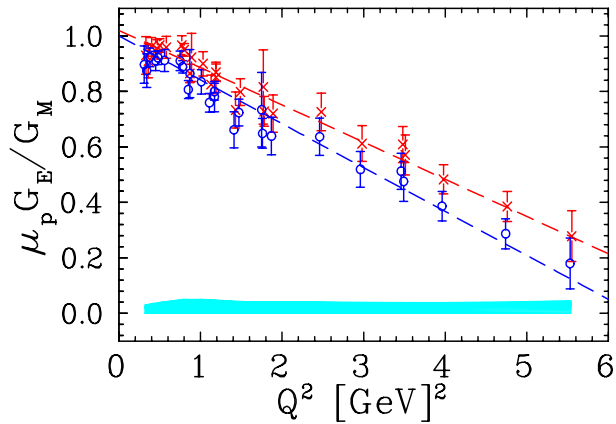}
\caption{(Color online) Polarization transfer measurements of $\mugegm$
as determined using the one photon exchange approximation (red $\times$) and
after applying the corrections based on the extraction of two-photon
contributions as described in the text (blue circles).  The error band at the
bottom shows the uncertainties associated with the two-photon corrections.
The corrected data are well fit by $\mugegm = 1 - 0.158 Q^2$ (bottom dashed
line).
\label{fig:newdata}}
\end{figure}

\begin{figure}[htb]
\includegraphics[height=5.5cm,width=8.0cm,angle=0]{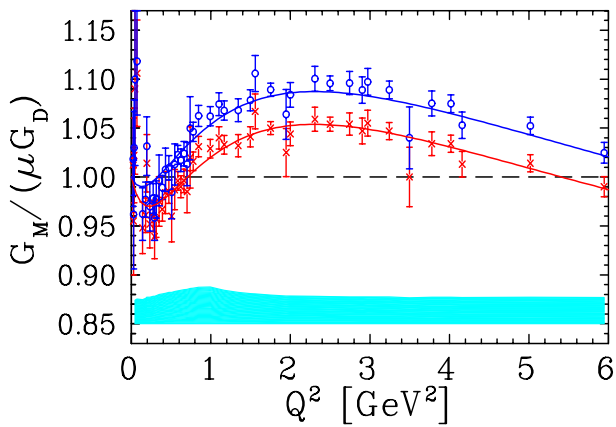}
\caption{(Color online) Rosenbluth extraction of $\gm / \mu_p \gd$
as determined using the one photon exchange approximation (red $\times$)
and after applying the corrections based on the extraction of two-photon
contributions as described in the text (blue circles).  The error band at the
bottom shows the uncertainties associated with the two-photon corrections.
The data are compared to fits from Ref.~\cite{arrington04a}, where 
the bottom curve (red) is the global analysis of the cross section data
while the top curve (blue) includes both cross section and polarization
transfer data, assuming an 6\%, linear, $\varepsilon$-dependent correction to
the cross section data.
\label{fig:newgm}}
\end{figure}

Having extracted the two-photon amplitudes, we can correct the polarization
transfer measurements to yield the true form factor ratio, $\gegm$.  We use
the fits to $\y2g$ and $\Delta\gm$ (Eqs.~\ref{eq:fit_y2g}
and~\ref{eq:fit_dgmgm}) to correct the polarization transfer measurements
according to Eq.~\ref{eq:r_poltrans}.  This yields a correction of
approximately 5\% at low $Q^2$ values, growing to 35\% at $Q^2=5.6$~GeV$^2$.
The fractional uncertainty in these amplitudes is $\sim$50\% for large $Q^2$
values (above 3-4 GeV$^2$), and increases to 100\% for low $Q^2$ values
($\approx 1$~GeV$^2$).  We parameterize the uncertainties in the extraction of
$\y2g$, $\Delta\gm/\gm$, and $\Delta\ge/ge$
(Eqs.~\ref{eq:dfit_y2g},~\ref{eq:dfit_dgmgm}, and ~\ref{eq:dfit_dgege}),
and use this to determine the uncertainty in the two-photon corrections we
apply to the polarization transfer ratio.  The dominant uncertainties in the
extraction of $\gegm$ are the experimental uncertainties in the polarization
transfer measurement (typically 3--15\%), the uncertainty in the extracted
values of $\y2g$ and $\Delta\gm$ (4--12\%), and the lack of knowledge of
$\Delta\ge$ (3--4\%). Figures~\ref{fig:newdata} and ~\ref{fig:newgm} show the
uncorrected and corrected values of $\gegm$ and $\gm$ with the experimental
uncertainties shown on the points and the uncertainties related to the
two-photon corrections shown in the error bar on the bottom.  A linear fit to
the corrected data yields $\mugegm = 1 - 0.158 Q^2$ (the uncorrected data yield
$\mugegm = 1 - 0.135 (Q^2-0.24)$~\cite{arrington03a}).

Next, we can use the low $\varepsilon$ cross sections, where $\y2g$ and
$\Delta\ge$ have little effect (Eq.\ref{eq:delta_sigma}), to extract
$\gmtilde$.  We take the limit as $\varepsilon \rightarrow 0$, as in the usual
L-T separation, and remove the two-photon contribution, $\Delta\gm$, as
determined from the above analysis to yield the corrected value for $\gm$. The
dominant uncertainties are the experimental cross section uncertainties
(1--2\% uncertainty in $\gm$) and the uncertainty in $\Delta\gm$ (1.5--3\%).
There is an additional uncertainty ($0.5\%$), coming from the uncertainty in
the large $\varepsilon$ ratio of positron to electron cross section, which is
only known to $\sim$1\% from the existing positron data.

In extracting the two-photon amplitudes and the uncertainties in
the corrected form factors we assumed that the two-photon amplitudes were
independent of $\varepsilon$. Any $\varepsilon$-dependence must be small
enough that it does not spoil the observed linearity of the reduced cross
sections. However, it could still be large enough to yield a noticeable
modification to the extracted value of $\gm$. If the amplitudes have
nonlinearities at low $\varepsilon$, then the value of $\gm$ will have an
additional correction. Most of the available Rosenbluth separations at large
$Q^2$ are limited to $\varepsilon \gtorder 0.2$, and therefore have to make a
significant extrapolation.  We can estimate the size of this uncertainty by
examining measurements of the linearity of the reduced cross sections that
have been performed as tests of the one-photon approximation.

A simple way to parameterize the limit on non-linearity is to fit the reduced
cross sections to a quadratic rather than a linear equation, $\sigma_r = P_0 (
1 + P_1 \varepsilon + P_2 \varepsilon^2$), and use the uncertainty on the
$\varepsilon^2$ coefficient, $P_2$, as an estimate of the possible nonlinear
terms.  The best linearity limits at high $Q^2$ come from the SLAC NE11
experiment~\cite{andivahis94}.  Their best measurement is at
$Q^2$=2.5~GeV$^2$, yields $P_2 = 0.0 \pm 0.11$. Similar limits ($\delta P_2 =
0.12$--$0.2$) are set by data at lower $Q^2$~\cite{berger71}. In
many cases, there are more data point at large $\varepsilon$ values,
increasing the uncertainty in the extrapolation to $\varepsilon=0$ beyond what
one would estimate from the simple quadratic fit. The simple estimate of
the allowed nonlinearities yields possible deviations of 3--4\%  in the
extraction of the cross section at $\varepsilon=0$, more for cases where
the linearity measurements are not as precise or where there is a larger
extrapolation to $\varepsilon = 0$.  We include a 4\% uncertainty on the cross
section (2\% uncertainty on $\gm$) due to the possible error in the
extrapolation to $\varepsilon=0$.

To obtain the corrected form factors, we assumed that the discrepancy is fully
explained by higher order radiative corrections, specifically two-photon
exchange and Coulomb distortion, and had to make some assumptions about the
$\varepsilon$-dependence.  To the extent that these are valid assumptions (or
good approximations), we obtain the correct two-photon amplitudes and
can obtain the corrected form factors. One way to test these assumptions is to
use the extracted amplitudes to predict the effect of these higher order terms
in comparisons of positron- and electron-proton scattering. While we
incorporate the $\varepsilon = 1$ constraint of the positron measurements into
the extraction, we have not included any other information on the
$\varepsilon$-dependence. Unfortunately, data at smaller $\varepsilon$ values
is limited to low $Q^2$, where the uncertainties in the two-photon amplitudes
are approaching 100\%. While we have to extrapolate to lower $Q^2$ to compare
to the positron data, we do reproduce the trend observed in the positron data.

\begin{figure}[htb]
\includegraphics[height=5.5cm,width=8.0cm,angle=0]{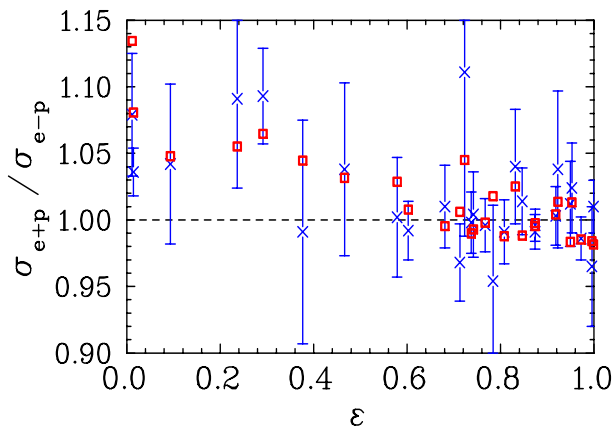}
\caption{(Color online) Ratio of positron-proton section to electron-proton
cross sections (blue $\times$), compared to the prediction from the extracted
values of $\y2g$ and $\Delta\gm/\gm$.
\label{fig:positron}}
\end{figure}

Figure~\ref{fig:positron} shows the ratio of positron to electron cross
section as a function of $\varepsilon$, along with the predictions based on
the fits to the two-photon amplitudes shown in Fig.~\ref{fig:amplitudes}. 
While the uncertainties are large, the data are in better agreement with the
prediction from the two-photon amplitudes: $\chi^2=18.2$ for 28 data points,
compared with $\chi^2$=23.9 if one assume $\sigma_{e+p} / \sigma_{e-p}=1$,
i.e. if the two-photon amplitudes are ignored.  While the positron data could
be included in the analysis to help constrain the two-photon amplitudes at low
$Q^2$, it is clear that the current data would not significantly modify the
values obtained using just the high-$\varepsilon$ constraint. Note that the
two-photon prediction shown here does not yield exactly unity for $\varepsilon
\rightarrow 1$. While the individual points for $\Delta\gm/\gm$ from
Fig.~\ref{fig:amplitudes} are determined by requiring that this ratio
approaches unity, the parameterizations of the two-photon amplitudes yield
slight deviations.

\section{Conclusions}

It is currently believed that the discrepancy comes from higher order
corrections to the Born approximation.  If this is true, and we assume a weak
$\varepsilon$-dependence of the two-photon amplitudes, we can use the
Rosenbluth, polarization transfer, and positron data to constrain the
two-photon exchange (and other order $\alpha_{EM}$) amplitudes.  The large
correction to Rosenbluth ratio allows extraction of $\y2g$, which then allows
determination of the (smaller) correction to the polarization transfer.  While
the $\y2g$ contribution to the polarization transfer ratio is as large as 35\%
for the largest $Q^2$ value, the overall trend of a roughly linear decrease in
$\mugegm$ with $Q^2$ remains.  Using the constraint from the two-photon (and
multi-photon) exchange, positron cross section measurements at large
$\varepsilon$, we also determine $\Delta\gm$.  While we cannot directly
constraint $\Delta\ge$, it has a relatively small effect on the extraction of
the form factors, as long as it is not much larger than the other two-photon
amplitudes.

Given these constraints on the two-photon amplitudes, we can correct
$\ge$ and $\gm$ for two-photon exchange effects, with 
additional uncertainties associated with these corrections. For $\gm$,
the uncertainty is dominated by possible $\varepsilon$-independence of
the amplitudes, which can lead to deviations from the linear extrapolation
to $\varepsilon=0$.  For $\ge$ (and $\gegm$), the uncertainties are dominated
by the large uncertainties in the $R_{L-T}$ data, which limit the precision
with which we can extract $\y2g$ and $\Delta\gm$.  We can use the two-photon
amplitudes we extract to provide corrected values of $\ge$ and $\gm$,
but with additional uncertainties related to these correction, yielding
final uncertainties that are 50-100\% larger than the experimental
uncertainties.

Additional positron data at low $\varepsilon$ and moderate $Q^2$ could be used
to test the assumption that the discrepancy is fully explained by two-photon
exchange, and give direct information on the $\varepsilon$-dependence.
Such measurements are planned at Novosibirsk~\cite{vepp_proposal}, and at
Jefferson Lab~\cite{e04116}. Additional Rosenbluth measurements, utilizing
the improved Rosenbluth technique of Ref.~\cite{arrington03e} can provide an
improved extraction of $R_{L-T}$ at moderate to large $Q^2$ values, as well as
better constraints on the linearity of the reduced cross section, and thus the
$\varepsilon$-dependence of the two-photon amplitudes. Additional constraints
on the $\varepsilon$-dependence of the two-photon amplitudes can be obtained
from measurements of the $\varepsilon$-dependence of the polarization
transfer~\cite{e04019}.  With such additional measurements, the two-photon
amplitudes can be constrained well enough that the proton form factors can be
extracted with uncertainties from the two-photon corrections that are
comparable to or smaller than the experimental uncertainties.

\begin{appendix}

\section{APPENDIX: Details of the two-photon amplitude extractions}

The following are the fits to the extracted values and uncertainties
for $\y2g$, $\Delta\gm/\gm$, and $\Delta\ge/\ge$:
\begin{eqnarray}
\label{eq:fit_y2g}
\y2g = 0.035 \cdot (1-\exp(-Q^2/1.45)) 			\\
\label{eq:fit_dgmgm}
\Delta\gm / \gm = (0.0124 Q^2 - 0.0445 \sqrt{Q^2}) \%		\\
\label{eq:fit_dgege}
\Delta\ge / \ge = 0 		\\
\label{eq:dfit_y2g}
\delta\y2g = (0.008 + 0.03 \exp(-Q^2/0.7) + 0.0015 Q^2) 	\\
\label{eq:dfit_dgmgm}
\frac{\delta(\Delta\gm/\gm)}{(\Delta\gm/\gm)} = \frac{\delta\y2g}{\y2g} \\
\label{eq:dfit_dgege}
\delta(\Delta\ge/\ge) = \max( ~\mid \y2g \mid~,~ 
\mid (\Delta\gm/\gm) \mid~)
\end{eqnarray}
with $Q^2$ in GeV$^2$, and with the relative uncertainty on the two-photon
corrections limited to 100\% at low $Q^2$ values.  At large $Q^2$ values
(above 3--4~GeV$^2$), the uncertainties in the two-photon amplitudes are
$\approx$40--50\%. Below $Q^2=1$~GeV$^2$, the parameterization of the
uncertainty on $\y2g$ becomes as large as the value itself, so the uncertainty
is taken to be 100\% in the analysis. Because $\Delta\gm/\gm$ is derived
directly from $\y2g$, the fractional uncertainty on $\Delta\gm$ is identical
to the fractional uncertainty on $\y2g$.  Finally, because we do not have any
data with which to constrain $\Delta\ge$, we assume that $\Delta\ge=0$, and
take the uncertainty to be the larger of the other two-photon amplitudes.

These fits and uncertainties are used to determine the correction to
$\gm$ as extracted from the cross section measurements, and the correction
to $\gegm$ as extracted from the polarization transfer data, as shown in
Fig.~\ref{fig:newdata}.  

Table~\ref{tab:newffs} gives the corrected values for $\gm$, determined by
taking the uncorrected value of $\gm$ (i.e. $\gmtilde$) from a global analysis
of the Rosenbluth data~\cite{arrington04a} and correcting for the extracted
value of $\Delta\gm$ (Eq.~\ref{eq:fit_dgmgm}). Additional two-photon
uncertainties come from the uncertainty in $\Delta\gm$ and a 2\% uncertainty
due to the possibility of non-linear terms that modify the extrapolation to
$\varepsilon=0$.

\setlength{\LTcapwidth}{8.7cm}
\begin{longtable}{@{\extracolsep{0cm}}p{3cm}p{3cm}}
\caption{Extracted values of $\gm$ after correcting for two-photon exchange
effects.  The form factor is given with respect to the dipole form,
$G_D = (1+Q^2/0.71)^{-2}$, with the experimental uncertainties listed first,
and the additional uncertainties related to two-photon effects listed second.
\label{tab:newffs}}
\\
$Q^2$ [GeV$^2$]	& $\gm/(\mu_pG_D)$	\\
\hline
0.005   & 0.751$\pm$0.424$\pm$0.016 \\
0.011   & 0.981$\pm$0.104$\pm$0.021 \\
0.015   & 1.000$\pm$0.074$\pm$0.021 \\
0.018   & 0.973$\pm$0.073$\pm$0.021 \\
0.022   & 1.018$\pm$0.049$\pm$0.022 \\
0.027   & 0.962$\pm$0.056$\pm$0.021 \\
0.034   & 1.030$\pm$0.049$\pm$0.022 \\
0.045   & 1.100$\pm$0.088$\pm$0.024 \\
0.072   & 1.118$\pm$0.055$\pm$0.025 \\
0.141   & 0.962$\pm$0.027$\pm$0.025 \\
0.179   & 0.977$\pm$0.018$\pm$0.026 \\
0.195   & 1.031$\pm$0.030$\pm$0.027 \\
0.234   & 0.970$\pm$0.025$\pm$0.027 \\
0.273   & 0.978$\pm$0.015$\pm$0.028 \\
0.292   & 0.958$\pm$0.023$\pm$0.028 \\
0.312   & 0.977$\pm$0.019$\pm$0.029 \\
0.350   & 0.996$\pm$0.026$\pm$0.030 \\
0.389   & 0.989$\pm$0.012$\pm$0.030 \\
0.428   & 1.007$\pm$0.022$\pm$0.031 \\
0.473   & 1.009$\pm$0.012$\pm$0.032 \\
0.507   & 0.984$\pm$0.027$\pm$0.032 \\
0.545   & 1.013$\pm$0.021$\pm$0.033 \\
0.584   & 1.023$\pm$0.010$\pm$0.034 \\
0.622   & 1.017$\pm$0.011$\pm$0.034 \\
0.663   & 1.024$\pm$0.012$\pm$0.035 \\
0.701   & 1.013$\pm$0.022$\pm$0.035 \\
0.740   & 1.050$\pm$0.032$\pm$0.036 \\
0.778   & 1.046$\pm$0.010$\pm$0.036 \\
0.846   & 1.062$\pm$0.011$\pm$0.037 \\
0.992   & 1.062$\pm$0.010$\pm$0.038 \\
1.102   & 1.074$\pm$0.012$\pm$0.035 \\
1.168   & 1.068$\pm$0.010$\pm$0.034 \\
1.344   & 1.068$\pm$0.012$\pm$0.032 \\
1.496   & 1.078$\pm$0.011$\pm$0.031 \\
1.557   & 1.106$\pm$0.018$\pm$0.031 \\
1.751   & 1.089$\pm$0.007$\pm$0.029 \\
1.947   & 1.064$\pm$0.025$\pm$0.028 \\
2.000   & 1.084$\pm$0.013$\pm$0.028 \\
2.308   & 1.100$\pm$0.013$\pm$0.028 \\
2.499   & 1.096$\pm$0.008$\pm$0.028 \\
2.743   & 1.096$\pm$0.012$\pm$0.028 \\
2.904   & 1.089$\pm$0.014$\pm$0.027 \\
2.972   & 1.097$\pm$0.014$\pm$0.028 \\
3.243   & 1.089$\pm$0.009$\pm$0.027 \\
3.497   & 1.040$\pm$0.032$\pm$0.027 \\
3.777   & 1.075$\pm$0.013$\pm$0.027 \\
4.018   & 1.075$\pm$0.009$\pm$0.027 \\
4.160   & 1.053$\pm$0.014$\pm$0.027 \\
5.017   & 1.052$\pm$0.009$\pm$0.027 \\
5.945   & 1.025$\pm$0.011$\pm$0.027 \\
7.037   & 0.989$\pm$0.015$\pm$0.026 \\
9.121   & 0.943$\pm$0.020$\pm$0.023 \\
\hline
\end{longtable}

\setlength{\LTcapwidth}{8.7cm}
\begin{longtable}{lccc}
\caption{Extracted form factor ratio $\gegm$ from polarization transfer
experiments and corresponding value of $\ge$ after applying the two-photon
corrections as described in the text.  In both cases, the experimental
uncertainty is listed first, and the additional uncertainty related to
the two-photon effects is listed second.
\label{tab:newratios}}
\\
Ref.	& $Q^2$		& $\mugegm$	& $\ge/G_D$	\\
	& [GeV$^2$]	& [corrected]	& [corrected]	\\
\hline
\cite{milbrath99}
& 0.380 & ~0.910$\pm$0.053$\pm$0.035~   & ~0.909$\pm$0.055$\pm$0.040~ \\
& 0.500 & ~0.969$\pm$0.053$\pm$0.043~   & ~0.979$\pm$0.055$\pm$0.048~ \\
\cite{pospischil01}
& 0.373 & ~0.961$\pm$0.054$\pm$0.035~   & ~0.959$\pm$0.056$\pm$0.040~ \\
& 0.401 & ~0.971$\pm$0.053$\pm$0.036~   & ~0.972$\pm$0.055$\pm$0.041~ \\
& 0.441 & ~0.895$\pm$0.051$\pm$0.036~   & ~0.899$\pm$0.053$\pm$0.041~ \\
\cite{jones00}
& 0.490 & ~0.921$\pm$0.025$\pm$0.039~   & ~0.930$\pm$0.028$\pm$0.044~ \\
& 0.790 & ~0.889$\pm$0.023$\pm$0.051~   & ~0.923$\pm$0.026$\pm$0.056~ \\
& 1.180 & ~0.800$\pm$0.030$\pm$0.047~   & ~0.854$\pm$0.033$\pm$0.053~ \\
& 1.480 & ~0.724$\pm$0.048$\pm$0.042~   & ~0.784$\pm$0.052$\pm$0.048~ \\
& 1.770 & ~0.649$\pm$0.054$\pm$0.040~   & ~0.708$\pm$0.059$\pm$0.046~ \\
& 1.880 & ~0.639$\pm$0.068$\pm$0.039~   & ~0.699$\pm$0.075$\pm$0.045~ \\
& 2.470 & ~0.637$\pm$0.068$\pm$0.039~   & ~0.699$\pm$0.075$\pm$0.045~ \\
& 2.970 & ~0.519$\pm$0.064$\pm$0.038~   & ~0.567$\pm$0.070$\pm$0.043~ \\
& 3.470 & ~0.512$\pm$0.065$\pm$0.039~   & ~0.555$\pm$0.071$\pm$0.044~ \\
\cite{gayou01}
& 0.320 & ~0.897$\pm$0.067$\pm$0.030~   & ~0.891$\pm$0.068$\pm$0.035~ \\
& 0.350 & ~0.875$\pm$0.061$\pm$0.031~   & ~0.871$\pm$0.062$\pm$0.036~ \\
& 0.390 & ~0.923$\pm$0.033$\pm$0.034~   & ~0.923$\pm$0.036$\pm$0.039~ \\
& 0.460 & ~0.910$\pm$0.035$\pm$0.037~   & ~0.916$\pm$0.037$\pm$0.042~ \\
& 0.570 & ~0.912$\pm$0.040$\pm$0.041~   & ~0.928$\pm$0.042$\pm$0.046~ \\
& 0.760 & ~0.910$\pm$0.035$\pm$0.047~   & ~0.943$\pm$0.038$\pm$0.053~ \\
& 0.860 & ~0.807$\pm$0.033$\pm$0.048~   & ~0.843$\pm$0.036$\pm$0.054~ \\
& 0.880 & ~0.864$\pm$0.087$\pm$0.049~   & ~0.904$\pm$0.091$\pm$0.055~ \\
& 1.020 & ~0.835$\pm$0.044$\pm$0.051~   & ~0.883$\pm$0.047$\pm$0.057~ \\
& 1.120 & ~0.759$\pm$0.034$\pm$0.047~   & ~0.808$\pm$0.037$\pm$0.053~ \\
& 1.180 & ~0.781$\pm$0.055$\pm$0.047~   & ~0.834$\pm$0.059$\pm$0.053~ \\
& 1.420 & ~0.662$\pm$0.065$\pm$0.041~   & ~0.715$\pm$0.071$\pm$0.047~ \\
& 1.760 & ~0.734$\pm$0.134$\pm$0.042~   & ~0.801$\pm$0.146$\pm$0.049~ \\
\cite{gayou02}
& 3.500 & ~0.475$\pm$0.072$\pm$0.038~   & ~0.515$\pm$0.078$\pm$0.043~ \\
& 3.970 & ~0.386$\pm$0.053$\pm$0.039~   & ~0.414$\pm$0.057$\pm$0.043~ \\
& 4.750 & ~0.287$\pm$0.054$\pm$0.041~   & ~0.302$\pm$0.057$\pm$0.044~ \\
& 5.540 & ~0.180$\pm$0.092$\pm$0.044~   & ~0.185$\pm$0.095$\pm$0.046~ \\
\end{longtable}

Given $\gegm$ and $\gm$, we can obtain $\ge$.  However, we have extracted
$\gegm$ from the polarization transfer measurements, and $\gm$ from the cross
section measurements, both corrected for the two-photon amplitudes.  Because
the uncertainty on $\gm$ is much smaller, we extract $\ge$ at the kinematics
of the polarization transfer measurements, using the corrected values of
$\gegm$ and a fit to the corrected values (and uncertainties) of $\gm$. 
Tables~\ref{tab:newffs} and~\ref{tab:newratios} give the corrected values for
$\gm$ and $\ge$, relative to the dipole form.

The corrected form factors are well described by the Polarization form factor
fit to $\gm$ from ref.~\cite{arrington04a} (top curve in
Fig.~\ref{fig:newgm}), and $\mugegm = 1 - 0.158 Q^2$ (bottom curve in
Fig.~\ref{fig:newdata}). The fit for $\gm$ from ref.~\cite{arrington04a} is
nearly identical to the best fit to the corrected $\gm$ data; the only
noticeable difference is that it is slightly lower (up to 1\%) for $Q^2$ values
of 2--4~GeV$^2$.  The fit by Brash, $\etal$~\cite{brash02}, is 1.5-2.5\% below
the corrected values of $\gm$ for $Q^2 \gtorder 1$~GeV$^2$.

\end{appendix}

\begin{acknowledgments}

The author wishes to thank Andrei Afanasev, Wally Melnitchouk, and Marc 
Vanderhaeghen for helpful discussions. This work was supported by the U. S.
Department of Energy, Office of Nuclear Physics, under contract
W-31-109-ENG-38.

\end{acknowledgments}

\bibliography{model_indep}

\end{document}